\documentclass[a4paper,fleqn,usenatbib]{mnras}

\usepackage{newtxtext,newtxmath}
\usepackage{graphicx}

\usepackage[T1]{fontenc}
\usepackage{ae,aecompl}

\usepackage{graphicx}	
\usepackage{amsmath}	

\newcommand{\pcm}{\,cm$^{-2}$}	
\newcommand{\cmmo}{\,cm$^{-1}$}	
\newcommand{\pcmc}{\,cm$^{-3}$} 

\title[Rotational excitation of H$_3$O$^+$ cations by $para-$H$_2$]{Rotational excitation of H$_3$O$^+$ cations by \textbf{\textit{para-}}H$_2$: improved collisional data at low temperatures}

\author[S. Demes et al.]{
S\'{a}ndor Demes$^{1}$, \thanks{E-mail: sandor.demes@univ-rennes1.fr} Fran\c{c}ois Lique$^{1}$, Alexandre Faure$^{2}$, Floris F. S. van der Tak$^{3}$, Claire Rist$^{2}$ and Pierre Hily-Blant$^{2}$
\\
$^{1}$ Univ Rennes, CNRS, IPR (Institut de Physique de Rennes) - UMR 6251, F-35000 Rennes, France\\ $^{2}$ IPAG, Universit\'{e} Grenoble Alpes \& CNRS, CS 40700, F-38058 Grenoble, France\\ $^{3}$ SRON Netherlands Institute for Space Research \& Kapteyn Astronomical Institute, University of Groningen, 9747 AD Groningen, The Netherlands\\
}

\date{Accepted XXX. Received YYY; in original form ZZZ}

\pubyear{2021}

\begin{document}
\label{firstpage}
\pagerange{\pageref{firstpage}--\pageref{lastpage}}
\maketitle

\begin{abstract}
The hydronium cation plays a crucial role in interstellar oxygen and water chemistry. While its spectroscopy was extensively investigated earlier, the collisional excitation of H$_3$O$^+$ is not well studied yet.

In this work we present state-to-state collisional data for rotational de-excitation of both $ortho$- and $para$-H$_3$O$^+$ due to $para$-H$_2$ impact. The cross sections are calculated within the close-coupling formalism using our recent, highly accurate rigid-rotor potential energy surface for this collision system. The corresponding thermal rate coefficients are calculated up to 100 K. For $para$-H$_3$O$^+$ the lowest 20 rotation-inversion states were considered in the calculations, while for $ortho$-H$_3$O$^+$ the lowest 11 states are involved (up to $j\leq5$), so all levels with rotational energy below 420 K (292 cm$^{-1}$) are studied.

In order to analyse the impact of the new collisional rates on the excitation of H$_3$O$^+$ in astrophysical environments radiative transfer calculations  are also provided. The most relevant emission lines from an astrophysical point of view are studied, taking into account the transitions at 307, 365, 389 and 396 GHz. We show that our new collisional data have a non-negligible impact (from a few percents up to about a factor of 3) on the brightness and excitation temperatures of H$_3$O$^+$, justifying the revision of the physical conditions in the appropriate astrophysical observations. The calculated rate coefficients allow one to recalculate the column density of hydronium in interstellar clouds, which can lead to a better understanding of interstellar water and oxygen chemistry.

\end{abstract}




\section{Introduction}\label{sec:Intro}

It was shown earlier by  \citet{Herbst1973,Phillips1992,Sternberg1995,VanderTak2008} that hydronium cation H$_3$O$^+$ is a very important chemical species in interstellar ion-molecule reactions schemes. It is present both in dense and diffuse interstellar medium (ISM) playing a very important role in oxygen and water chemistry \citep{Sternberg1995,Gonzalez2013,Goicoechea2001,Gerin2010,VanDishoeck2013}. According to \citet{Hollenbach2012} the formation of H$_3$O$^+$ proceeds via two different ways in low-temperature diffuse and dense molecular clouds (see also \citet{VanDishoeck2013}). In diffuse clouds as well as near the surfaces of molecular clouds the following reactions take place:
\begin{equation}
    (a) \> \mathrm{OH}^+ + \mathrm{H}_2 \rightarrow \mathrm{H}_2\mathrm{O}^+ + \mathrm{H} , \\
    (b) \> \mathrm{H}_2\mathrm{O}^+ + \mathrm{H}_2 \rightarrow \mathrm{H}_3\mathrm{O}^+ + \mathrm{H} .
	\label{eq:OHp}
\end{equation}
Here the OH$^+$ cations are formed by an "atomic" route via the interaction of O$^+$ ions with H$_2$, where the oxygen ions are produced in H$^+ +$  O reaction.

Deeper in the opaque interiors of molecular clouds the formation of H$_3$O$^+$ can take place also via reactions $(a)$ and $(b)$ defined in Eq.~(\ref{eq:OHp}). In this region however both OH$^+$ and H$_2$O$^+$ cations can be produced directly from H$_3^+$ through interaction with oxygen atoms, so reaction $(a)$ could be skipped.
H$^+$ and H$_3^+$ ions needed for H$_3$O$^+$ formation are provided by cosmic-ray ionization of H$_2$. In our Galaxy, this process proceeds at a rate of $10^{-17}-10^{-15}$ $\mathrm{s}^{-1}$, depending on cloud density and Galactocentric radius \citep{VanDerTak2000,Indriolo2015}. The relative abundances of H$_2$O and H$_3$O$^+$ have been used to constrain the ionization rates of molecular clouds in external galaxies \citep{VanDerTak2016}.

The dissociative recombination of H$_3$O$^+$ leads to the formation of neutral OH and H$_2$O products, where the hydroxyl radical can react with oxygen atoms afterward in order to form O$_2$ molecules \citep{VejbyChristensen1997,Jensen2000,Zhaunerchyk2009,SYu2009}. H$_3$O$^+$ is not only the backbone of interstellar oxygen chemistry, but it is also one of the most important cations in all branches of chemistry in general \citep{Yu2016,Mann2013}, which is intensively studied in many aspects. As we mentioned in our recent work (\citet{Demes2020}, referred as Paper I hereafter), finding the abundance of H$_3$O$^+$ cations can be used as an indirect way of determining the abundance of interstellar water \citep{Phillips1992}. In order to understand the role of hydronium cations in astrophysical media however the features of both its radiative and collisional excitation should be studied. Numerous high-resolution spectroscopy measurements were carried out for H$_3$O$^+$, which were collected by \citet{SYu2009}, while \citet{Yurchenko2020} theoretically studied its ro-vibrational levels recently. The electron-impact excitation of H$_3$O$^+$ was studied by \citet{Faure2003}, who combined the {\bf R}-matrix theory with the adiabatic-nuclei-rotation approximation. Unfortunately there are only limited works devoted to its collisions with interstellar atoms and molecules. For example, for rotational excitation of H$_3$O$^+$ by molecular hydrogen (which is the most abundant molecule in the ISM) only approximated, scaled data are available in the literature \citep{Offer1992}, which are only accurate to order of magnitude (see the review by \citet{VanDerTak2020} for a discussion of the accuracy of scaled collision data). So reliable and precise rate coefficients are obviously needed for describing the collisional excitation of H$_3$O$^+$ in order to interpret the astrophysical observations in the ISM.

In the first rotational excitation study by \citet{Offer1992} the H$_3$O$^+ - $ H$_2$ collision was modeled with a scaled interaction potential of the isoelectronic NH$_3$ molecule (in collision with H$_2$), taking into account its electronic and geometrical properties along with the large ``umbrella'' inversion splitting constant. Both the {\it para}($p$)- and {\it ortho}($o$)-H$_3$O$^+$ nuclear spin isomers were studied in collision with ground-state {\it para}- and {\it ortho}-H$_2$. In order to improve the model potential energy surface (PES) based on NH$_3$-H$_2$ interaction, the authors added a long-range correction calculated by second-order perturbation theory, fitted by analytical functions. However, even with this correction included the model cannot correctly describe the ionic nature of H$_3$O$^+$, so the cross section data obtained by \citet{Offer1992} should be improved and recalculated using a more accurate PES. It is also worth noting that the authors provide converged state-to-state data only for the $p$-H$_3$O$^+ -$ $p$-H$_2$ collision for a single collision energy ($101$ \cmmo), while for the $o$-H$_3$O$^+ -$ $p$-H$_2$ and $p$-H$_3$O$^+ -$ $o$-H$_2$ interactions they only performed benchmark calculations with a very limited number of partial waves $J=0-2$ (where $J$ is the total angular momentum).

The collisional excitation of H$_3$O$^+$ by He atoms (as a template for H$_2$) was recently studied by \citet{ElHanini2019}. It is worth to mention however that He is a factor of 5 less abundant than H$_2$. Rotational excitation cross sections were computed by means of the close coupling or coupled channel (CC) method by the authors up to 500 \cmmo collision energies in case of {\it o}-H$_3$O$^+$, and up to 300 \cmmo in case of {\it p}-H$_3$O$^+$, while the thermal rate coefficients were calculated up to 50 K. They used a 3-dimensional PES with fixed intramolecular distances for both colliding partners, calculated by the coupled-cluster theory at the level of single, double, and perturbative corrections for triple excitations [CCSD(T)] and with a non-standard valence quadruple-zeta (AVQZ) basis set. The inversion splitting effects were neglected in the work, which is questionable in case of the H$_3$O$^+$ cation because of its rather large ($\sim$ 55 \cmmo) inversion splitting constant \citep{SYu2009}. The global well depth of the H$_3$O$^+$ -- He system, as it was calculated by the authors, taking into account the basis set superposition error (BSSE) corrections, is about 354.53 \cmmo. It is worth to mention that, as it was shown by \citet{Roueff2013}, significant differences could be observed between the rate coefficients for collisions with He and H$_2$. According to this, new, reliable rate coefficients are obviously needed for the collision of H$_3$O$^+$ with H$_2$ in order to more precisely estimate its abundance from emission spectral lines observed in the ISM. Without such rates the molecular abundances could be only approximated, assuming local thermodynamic equilibrium (LTE), which is usually not a good approximation for typical interstellar conditions \citep{Roueff2013}.

In this paper, we provide new collisional cross section and thermal rate coefficient data for the interaction of H$_3$O$^+$ with H$_2$ under ISM conditions, based on our new, accurate 5-dimensional PES \citep{Demes2020}. The rotational (non-spherical) structure of $p$-H$_2$ was taken into consideration.A detailed comparison with the available data for H$_3$O$^+$ -- He \citep{ElHanini2019} and NH$_3$ -- H$_2$ \citep{Bouhafs2017} collisions is presented. As a first astrophysical application, in order to illustrate the influence of our new rate coefficients on astrophysical modelling, we also perform radiative transfer calculations to analyze the collisional excitation of H$_3$O$^+$ in interstellar molecular clouds.

The paper is organized as follows: in Section~\ref{sec:ModMeth} the details of the PES and the scattering calculations are presented. We report state-to-state cross sections and rate coefficients for the rotational de-excitation of H$_3$O$^+$ by H$_2$ in Section~\ref{sec:Results}, where in subsection~\ref{sec:Excit} we also present the results of radiative transfer calculation. Our concluding remarks are drawn in Section~\ref{sec:Conclusions}.

\section{Model and method}\label{sec:ModMeth}

\subsection{Potential energy surface}\label{sec:PES}
We have used our recent 5D rigid-rotor potential energy surface \citep{Demes2020} to model the collisional dynamics of the H$_3$O$^+$–H$_2$ system. The PES was calculated using the explicitly correlated coupled-cluster theory at the level of singles and doubles with perturbative corrections for triple excitations [CCSD(T)-F12] with the moderate-size augmented correlation-consistent valence triple zeta (aug-cc-pVTZ) basis set and applying the counterpoise procedure of \citet{Boys1970} in order to eliminate the effects of basis set superposition error (see \citet{Demes2020} for full details).

We used a Jacobi coordinate system to define the geometry of the H$_3$O$^+$ -- H$_2$ system. The center of the coordinate system is chosen to be in the center of mass (c.o.m.) of the H$_3$O$^+$ cation (i.e. molecular frame representation). A single radial parameter $R$ defines the intermolecular distance between the c.o.m. of H$_3$O$^+$ and H$_2$, $\theta$ and $\phi$ spherical angles characterize the angular position of H$_2$ relative to the center of the coordinate system, while two additional angles define the orientation of the H$_2$ molecule. We used the rigid rotor approximation in order to reduce the dimensionality. It should be noted that the inversion frequency of H$_3$O$^+$ is much larger than that of NH$_3$, however, as \citet{Offer1992} have shown the explicit inclusion of the inversion motion has only a small impact on the rotational cross sections, so that the rigid-rotor approximation is valid and reasonable for our case. For the H$_2$ molecule we applied a bond length of $r_\mathrm{H-H} = 1.44874$ bohr, calculated by \citet{Bubin2003}, while for the H$_3$O$^+$ cation the experimental structural data of \citet{Tang1999} was used: $r_\mathrm{O-H} = 1.8406$ bohr for the bond lengths and $\alpha_\mathrm{H-O-H} = 113.6^{\circ}$ for the bond angles. At each intermolecular distance $R$, the interaction energy was fitted using a least square standard linear method. All significant analytical terms were selected iteratively, using a Monte-Carlo error estimator described by \cite{rist2012}, resulting in a final set of 208 expansion functions with anisotropies up to $l_1=16$ and $l_2=4$ (the $l_1$ and $l_2$ indices refer to the tensor ranks of the angular dependence of the H$_3$O$^+$ and H$_2$ orientations, respectively). The root mean square (rms) residual was found to be lower than 1~\cmmo in the long-range region of the PES and also in the potential well region. The rms error on the expansion coefficients was also found to be smaller than 1 \cmmo in these regions of the interaction potential. The interpolation of the analytical coefficients was finally performed using a cubic spline radial interpolation scheme over the whole intermolecular distance range ($R=4-30$ bohr), and it was smoothly connected to standard extrapolations using the switch function applied earlier by \citet{Valiron2008}.

Until otherwise noted, we used the following units throughout this paper: atomic units (a.u.) for distances (1~a.u. = 1~bohr $\approx 5.29177 \times 10^{-9}$ cm), and wavenumbers (\cmmo) for energies (1~\cmmo $\approx 1.4388$ K).

\subsection{Scattering calculations}\label{sec:ScatCalc}

We have calculated the state-to-state rotational de-excitation cross sections ($\sigma$) and thermal rate coefficients ($k(T)$) for the collision of both {\it ortho-} and {\it para-}H$_3$O$^+$ with {\it para-}H$_2$. We determined the $\sigma$ up to 800 cm$^{-1}$ total energies by the close-coupling approach using the \texttt{HIBRIDON} scattering code \citep{Manolopoulos86,Millard87}. Similar calculations were performed for the isoelectronic NH$_3$--H$_2$ collision system (which is also a ``symmetric top molecule -- diatom'' collision system) in a broad kinetic energy range by \citet{Bouhafs2017}. The rotational states involved in our calculations are denoted by $j_k^{\epsilon}$, where $j$ is the total angular momentum of the H$_3$O$^+$ cation, $k$ is its projection on the $C_3$ rotational axis, and $\epsilon=\pm$ is an inversion symmetry index (for more details, see \cite{Rist1993}). It is worth noting that the {\it ortho-}H$_3$O$^+$ is characterised with $k=3n$ quantum numbers (where $n=0,1,2,\dots$), while all other $k$ quantum numbers (e.g. $k=1,2,4,5,\dots$) define the H$_3$O$^+$ with {\it para} nuclear spin symmetry.

In case of the hydronium cation, we used the most recent experimental rotational constants from \citet{SYu2009}: $B=11.15458$ cm$^{-1}$ and $C=6.19102$ cm$^{-1}$. The ground-state inversion splitting constant of H$_3$O$^+$ measured by \citet{SYu2009} is 55.34997 cm$^{-1}$, which is nearly two orders of magnitude larger than that of NH$_3$ (0.7903 cm$^{-1}$). In case of the H$_2$ molecule we considered the $B_{\mathrm{H}_2}=59.3801$ cm$^{-1}$ rotational constant, in accordance with \citet{Bouhafs2017}. The reduced mass of the H$_3$O$^+$--H$_2$ system is found to be 1.82273 amu. Just as in Paper I we only used a reduced 55-term expansion of our analytical PES in current scattering calculations, including anisotropies up to $l_1\leq 6$ and $l_2\leq 2$. The implementation of the full, 208-term angular basis into the scattering code would be technically difficult and would make the close-coupling calculations unnecessarily heavy and computationally too expensive. However, in order to study the effect of the size of the analytical potential on the cross sections we performed benchmark calculations also with a 39-term angular basis, in which all terms with $l_1=6$ were removed. The results show that the 55-term angular basis we used in the calculations is enough to ensure a convergence always better than $10-20\%$ for any particular energy and transition considered, and generally keep the accuracy on the level of a few percents.

We calculated the inelastic cross sections for transitions between rotational levels with internal energy $\leq420$ K ($292$ cm$^{-1}$). The complete set of these rotational levels for both {\it ortho-} and {\it para-}H$_3$O$^+$ are shown in Table~\ref{tab:rot-levels}. The basis set size ($j_\mathrm{max}$) along with the maximum total angular momentum ($J_\mathrm{tot}$) were selected depending on the collision energy. Their particular values were found following preliminary convergence test calculations with a convergence threshold criteria of 1\% and 0.01\% mean deviation, respectively. In order to adequately describe the resonances in the cross sections, a very small energy step size ($E_\mathrm{step}$) was chosen at low collision energies (0.1 \cmmo), which was then gradually increased (until 5 \cmmo at high energies). The values of $j_\mathrm{max}$, $J_\mathrm{tot}$ and $E_\mathrm{step}$ parameters used in the calculations for the particular collision energy intervals (from $E_\mathrm{init}$ to $E_\mathrm{fin}$) are listed in Table~\ref{tab:conv-param}. It is worth to mention, that the $j_\mathrm{max}$ parameter refers only to the rotational basis used for the H$_3$O$^+$ cation. For {\it para}-H$_2$ the two lowest even rotational states were taken into account in the scattering calculations, i.e. $j_{\mathrm{H}_2}$ = 0, 2.

\begin{table*}
	\centering
    \caption{List of the rotational energy levels for {\it ortho-} and {\it para-}H$_3$O$^+$, which were considered in collisional studies of the present work. The corresponding energies are taken from JPL database \citep[Species Tag: 19004]{Pickett2010}.}
    \label{tab:rot-levels}
    \begin{tabular}{ccc|ccc}
        \hline
        \multicolumn{3}{c}{\textbf{\textit{para-}H$_3$O$^+$}} &  \multicolumn{3}{c}{\textbf{\textit{ortho-}H$_3$O$^+$}}\\
        \hline
        state  & rotational  & rotational (internal) & state  & rotational  & rotational (internal) \\
        label & state $j_k^{\epsilon}$  & energy [\cmmo] & label & state $j_k^{\epsilon}$ & energy [\cmmo] \\
        \hline
        (1) & $1_1^+$	&   0.000 & (1) & $1_0^+$ &   5.101	 \\
        (2) & $2_2^+$	&  29.697 & (2) & $0_0^-$ &  37.947	 \\
        (3) & $2_1^+$	&  44.986 & (3) & $3_3^+$ &  71.681	 \\
        (4) & $1_1^-$	&  55.233 & (4) & $2_0^-$	& 104.239	 \\
        (5) & $2_2^-$	&  84.977 & (5) & $3_0^+$	& 117.457	 \\
        (6) & $3_2^+$	&  97.145 & (6) & $3_3^-$	& 127.172	 \\
        (7) & $2_1^-$	&  99.427 & (7) & $4_3^+$ & 161.573	 \\
        (8) & $3_1^+$	& 112.384 & (8) & $4_3^-$	& 215.486	 \\
        (9) & $4_4^+$	& 125.939 & (9) & $4_0^-$	& 258.636	 \\
        (10) & $3_2^-$	& 151.241 & (10) & $6_6^+$ & 271.201	 \\
        (11) & $3_1^-$	& 165.657 & (11) & $5_3^+$ & 273.697	 \\
        (12) & $4_4^-$	& 181.807					 \\
        (13) & $4_2^+$	& 186.926					 \\
        (14) & $5_5^+$	& 192.453					 \\
        (15) & $4_1^+$	& 202.099					 \\
        (16) & $5_4^+$	& 238.256					 \\
        (17) & $4_2^-$	& 239.479					 \\
        (18) & $5_5^-$	& 248.863					 \\
        (19) & $4_1^-$	& 253.850					 \\
        (20) & $5_4^-$	& 292.146					 \\
        \hline
    \end{tabular}
\end{table*}

The calculated cross sections allow to determine the rate coefficients up to $100$ K kinetic temperatures by integrating over a Maxwell-Boltzmann distribution of relative velocities \citep{LePicard2019}: 

\begin{equation}
    k_{\mathrm{i} \rightarrow \mathrm{f}}(T) = \left(\frac{8}{\pi\mu k_\mathrm{B}^3 T^3}\right)^\frac{1}{2} \int_{0}^{\infty} \sigma_{\mathrm{i} \rightarrow \mathrm{f}} E_\mathrm{c} e^{-\frac{E_\mathrm{c}}{k_\mathrm{B}T}} dE_\mathrm{c} ,
    \label{eq:rates}
\end{equation}
where $E_\mathrm{c}$ is the collision energy, $\sigma_{\mathrm{i} \rightarrow \mathrm{f}}$ is the cross section for transition from the initial state (i) to the final state (f), $\mu$ is the reduced mass of the system and $k_\mathrm{B}$ is the Boltzmann constant.

\begin{table*}
	\centering
    \caption{A full list of the basis set size ($j_\mathrm{max}$) and maximum total angular momentum ($J_\mathrm{tot}$) parameters, which were found by systematic convergence tests for the particular total energy intervals. The step size for the energies ($E_\mathrm{step}$) used in the computations are also listed.}
    \label{tab:conv-param}
    \begin{tabular}{cccccc|ccccc}
        \hline
        \multicolumn{5}{c}{\textbf{\textit{ortho-}H$_3$O$^+ -$ \textit{para-}H$_2$ collision}} & &  \multicolumn{5}{c}{\textbf{\textit{para-}H$_3$O$^+ -$ \textit{para-}H$_2$ collision}} \\
        \hline
        $E_\mathrm{init}$  & $E_\mathrm{fin}$  & $E_\mathrm{step}$ & $j_\mathrm{max}$ & $J_\mathrm{tot}$ &  & $E_\mathrm{init}$  & $E_\mathrm{fin}$  & $E_\mathrm{step}$ & $j_\mathrm{max}$ & $J_\mathrm{tot}$ \\
        $[$cm$^{-1}]$ & $[$\cmmo$]$  & $[$\cmmo$]$ &  & & & $[$cm$^{-1}]$ & $[$\cmmo$]$  & $[$\cmmo$]$ &  &  \\
        \hline
        55.4 & 99.9 & 0.1 & 9 & 23 & &47.1 & 74.9 & 0.1 & 17 & 18 \\
        100 & 149.9 & 0.1 & 9 & 30 & &75 & 99.9 & 0.1 & 17 & 22 \\
        150 & 174.5 & 0.5 & 9 & 34 & &100 & 124.9 & 0.1 & 17 & 26 \\
        175 & 199.5 & 0.5 & 9 & 35 & &125 & 149.9 & 0.1 & 17 & 30 \\
        200 & 224.5 & 0.5 & 10 & 42 & &150 & 174.5 & 0.5 & 17 & 34 \\
        225 & 249.5 & 0.5 & 10 & 43 & &175 & 199.5 & 0.5 & 17 & 38 \\
        250 & 349.5 & 0.5 & 10 & 50 & &200 & 224 & 1 & 17 & 39 \\
        350 & 374.5 & 0.5 & 10 & 51 & &225 & 249 & 1 & 17 & 43 \\
        375 & 399.5 & 0.5 & 10 & 55 & &250 & 274 & 1 & 17 & 44 \\
        400 & 424.5 & 0.5 & 10 & 56 & &275 & 299 & 1 & 17 & 48 \\
        425 & 449.5 & 0.5 & 12 & 57 & &300 & 324 & 1 & 17 & 49 \\
        450 & 474.5 & 0.5 & 12 & 58 & &325 & 374 & 1 & 17 & 51 \\
        475 & 499.5 & 0.5 & 12 & 60 & &375 & 399 & 1 & 17 & 52 \\
        500 & 508 & 2 & 12 & 73 & &400 & 424 & 1 & 17 & 53 \\
        510 & 522.5 & 2.5 & 12 & 73 & &425 & 449 & 1 & 17 & 54 \\
        525 & 547.5 & 2.5 & 12 & 74 & &450 & 474 & 1 & 17 & 55 \\
        550 & 570 & 5 & 12 & 74 & &475 & 499 & 1 & 17 & 56 \\
        575 & 620 & 5 & 12 & 79 & &500 & 518.75 & 1.25 & 17 & 58 \\
        625 & 645 & 5 & 12 & 81 & &520 & 547.5 & 2.5 & 17 & 58 \\
        650 & 670 & 5 & 12 & 82 & &550 & 570 & 5 & 17 & 59 \\
        675 & 695 & 5 & 12 & 83 & &575 & 620 & 5 & 17 & 60 \\
        700 & 720 & 5 & 13 & 84 & &625 & 695 & 5 & 16 & 62 \\
        725 & 745 & 5 & 13 & 85 & &700 & 745 & 5 & 16 & 63 \\
        750 & 770 & 5 & 13 & 87 & &750 & 800 & 5 & 16 & 65 \\
        775 & 800 & 5 & 13 & 93 &  &  &  &  &  & \\
        \hline
    \end{tabular}
\end{table*}

\subsection{Radiative transfer calculations}\label{sec:RadTRCalc}

Using the new set of collisional rate coefficients, radiative transfer calculations were performed in order to determine the $T_{\mathrm{R}}$ radiance and $T_{\mathrm{exc}}$ excitation temperatures as well as the $\tau$ opacity (optical depth) parameters for radiative transitions of {\it ortho-} and {\it para-}H$_3$O$^+$ under specific interstellar physical conditions. The calculations were performed using the latest version of the \texttt{RADEX} non-LTE radiative transfer computer program \citep{VanDerTak2007}. In order to solve the radiative transfer equation, the uniform sphere escape probability method has been applied. In the calculations the physical conditions were either systematically varied or constrained. For example, the {\it para-}H$_2$ molecular densities were considered in the interval of $n(\mathrm{H}_2) = 10^{3}-10^{12}$ \pcmc , the molecular column density $N$ was varied from $1\times10^{12}$ up to $1\times10^{15}$~\pcm , while the calculated rate coefficients allowed one to explore a kinetic energy range up to $T_{\mathrm{kin}} = 100$ K. The background radiation temperature and the FWHM line width parameters were kept fixed in all calculations at $T_\mathrm{bg} = 2.7$~K and $\Delta V = 10.0$~km/s, respectively.

Two sets of radiative transfer calculations were provided. First, we used the available rate coefficients from the \texttt{LAMDA} database \citep{Schoier2005}. This database contains scaled {\it ortho-} and {\it para-}NH$_3$ rate coefficients in collision with H$_2$ at 100 K kinetic temperature to estimate the H$_3$O$^+$-H$_2$ collisional rates. The \texttt{LAMDA} database provides such rates for 36 de-excitation transitions in case of {\it ortho-} and 91 transitions in case of {\it para-}H$_3$O$^+$.

In the second set of radiative transfer calculations we used the new rate coefficients, which were calculated according to the method discussed in subsection~\ref{sec:ScatCalc}. We have explicitly taken into account the rate coefficients between 10 and 100 K kinetic temperatures for 55 transitions in case of {\it ortho-} and 180 transitions in case of {\it para-}H$_3$O$^+$. For both sets of calculations the rotational energies along with the radiative transition frequencies and Einstein $A$-coefficients were used as it is proposed in JPL database \citep[Species Tag: 19004]{Pickett2010}. In total 11 rotational states (up to the $j_k^{\epsilon} = 5_{3}^+$ state) were included in the analysis of {\it ortho-}H$_3$O$^+$, while in case of {\it para-}H$_3$O$^+$ we considered the 20 lowest states (including the $j_k^{\epsilon} = 5_{4}^-$ state as the highest, see Table.~\ref{tab:rot-levels} for more details).

\section{Results and discussion}\label{sec:Results}
\subsection{Rotational de-excitation cross sections}\label{sec:CSs}

\begin{figure}
    \centering
		\includegraphics[width=\columnwidth]{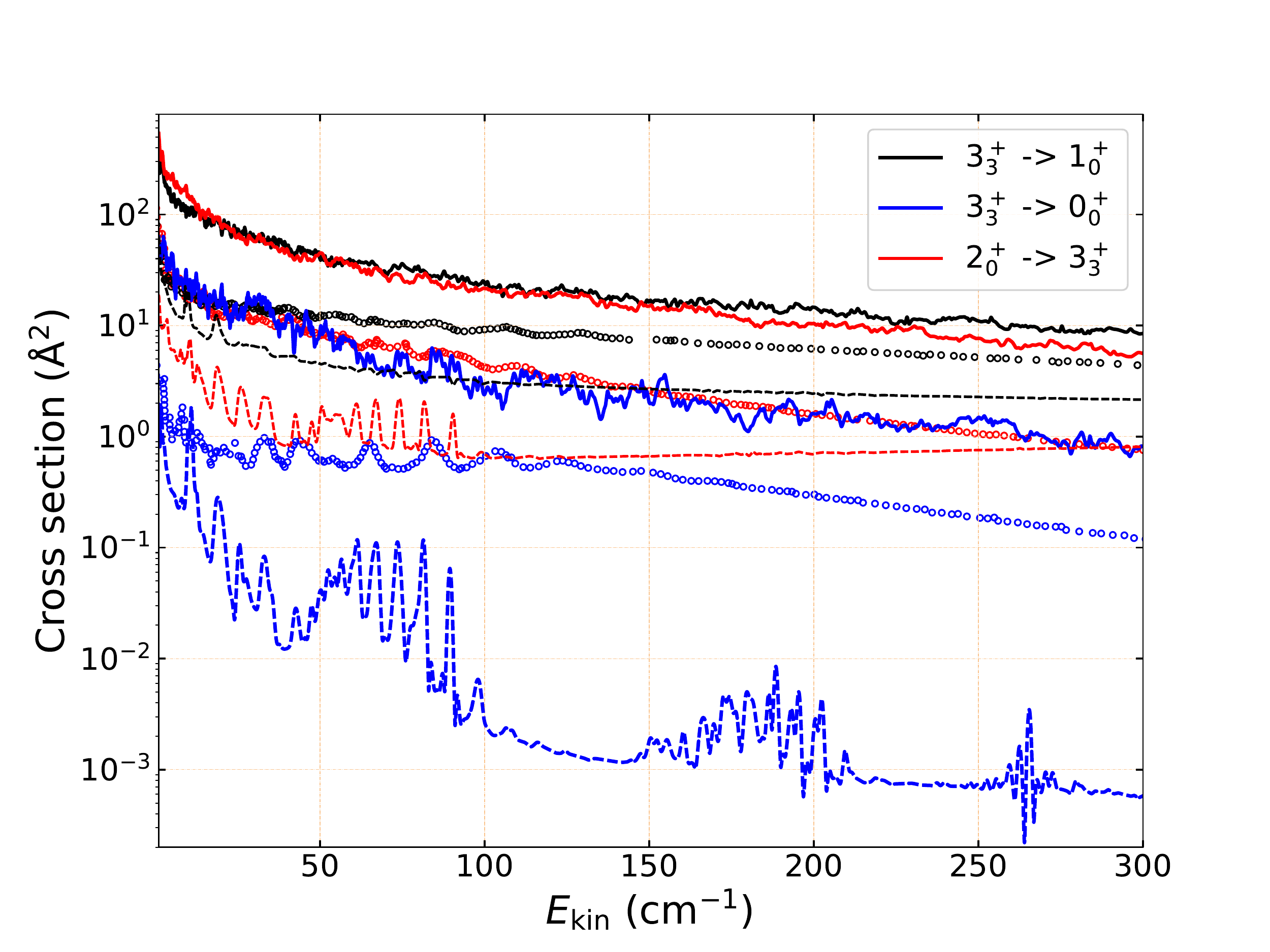}
    \caption{Rotational de-excitation cross sections for some selected transitions in collision of $o-$H$_3$O$^+$ with $p-$H$_2$. Our results (solid lines) are compared with the corresponding data of \citet{Bouhafs2017} ($o-$NH$_3$ -- $p-$H$_2$, dashed lines) and \citet{ElHanini2019} ($o-$H$_3$O$^+$ -- He, circles). Identical colours indicate identical transitions.}
    \label{fig:CSop}
\end{figure}

\begin{figure}
    \centering
		\includegraphics[width=\columnwidth]{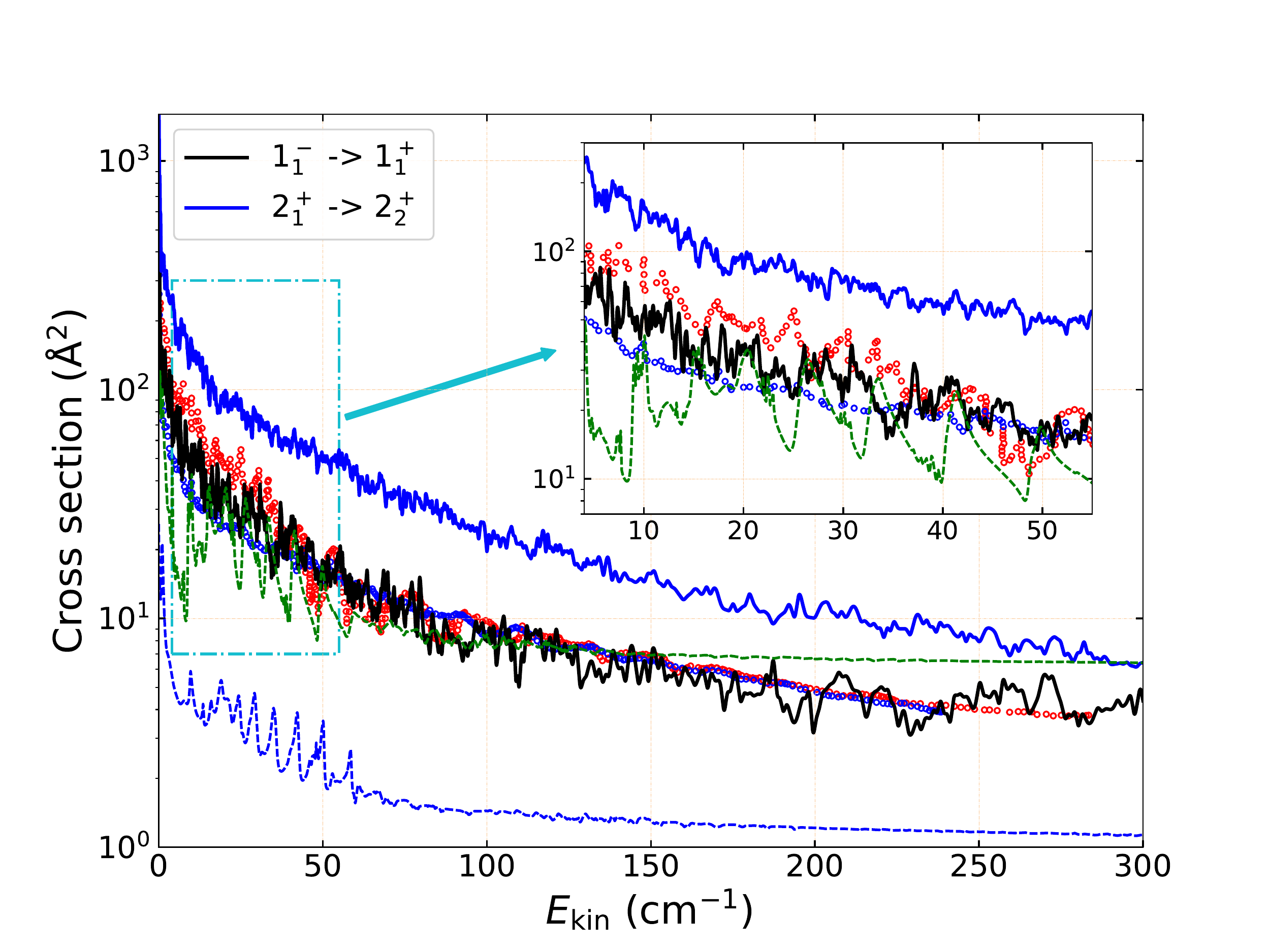}
    \caption{Rotational de-excitation cross sections for some selected transitions in collision of $p-$H$_3$O$^+$ with $p-$H$_2$. Our results (solid lines) are compared with the corresponding data of \citet{Bouhafs2017} ($p-$NH$_3$ -- $p-$H$_2$, dashed lines) and \citet{ElHanini2019} ($p-$H$_3$O$^+$ -- He, circles). The blue lines and circles are for the $2_1^+ \rightarrow 2_2^+$ transition, while both the green dashed line and the red circles show the data for $1_1^- \rightarrow 1_1^+$ transition.}
    \label{fig:CSpp}
\end{figure}

Figs.~\ref{fig:CSop} and~\ref{fig:CSpp} show the collisional energy dependence of some selected de-excitation cross sections for $o-$H$_3$O$^+$ and $p-$H$_3$O$^+$ collision with $p$-H$_2$, respectively. We compare our data with the corresponding cross sections for H$_3$O$^+$ -- He \citep{ElHanini2019} and NH$_3$ -- H$_2$ \citep{Bouhafs2017} collisions. The reason why we perform a comparative analysis also with ammonia data is because NH$_3$ was assumed to be a template for the collisions with H$_3$O$^+$ \citep[see for example][for details]{Offer1992}. It is worth to mention however, that the labeling of the rotational levels in both of the referred works is somewhat different from those we calculated for the hydronium cation (see Table.~\ref{tab:rot-levels}). The source of such differences is obvious in comparison with NH$_3$, however the incorrect structure of rotational levels for H$_3$O$^+$ is because the large inversion splitting constant of the cation is neglected in the calculations of \citet{ElHanini2019}.

It is worth to mention that \citet{Offer1992} only calculated excitation cross sections and only for one single energy both for $para$- and $ortho-$H$_3$O$^+$ collision with H$_2$. These cross sections were calculated in the low-energy region (at 101 and 145~\cmmo, respectively), where the resonances are very dominant, and their structure  and amplitude strongly depend on the quality of the potential used for the calculations. We compared the state-to-state cross sections with the corresponding data of \citet{Offer1992}, and we found that our results usually highly overestimates them (up to 1 or even 2 orders of magnitude). It is important to mention also that in order to make a comparison of the de-excitation cross sections directly in Figs.~\ref{fig:CSop} and~\ref{fig:CSpp} an analytical conversion is needed. This conversion involves the $k_j^2$ (wavevector) parameter, which linearly depends on the difference between the total collision energy and the rotational energy of the particular states involved in the transition. Since the quality of the conversion is sensitive to the rotational level energy (especially in case of higher states), which are not listed precisely in the work of \citet{Offer1992} (see Fig. 3 therein), it is not possible to guarantee a good accuracy for the conversion. Because of the mentioned reasons we decided not to include the results of \citet{Offer1992} in our global analysis.

In the case of $o-$H$_3$O$^+$ collision with $p$-H$_2$ we selected some transitions with $\Delta j = 1-3$ and $\Delta k = 3$. As one can see there is a significant difference between our results and the cross sections of both \citet{ElHanini2019} and \citet{Bouhafs2017} in the whole kinetic energy interval. Typically our cross sections are about an order of magnitude larger than the corresponding data from the literature. The relative difference is the smallest in case of the $3_3^+ \rightarrow 1_0^+$ transition, and somewhat decreases with increasing energy. For other transitions the difference is relatively constant and does not depend strongly on energy, except in the case of the $2_0^+ \rightarrow 3_3^+$ transition in comparison with the NH$_3$ data. Significantly there is no constant linear scaling between our data and those of \citet{ElHanini2019} or \citet{Bouhafs2017}, which confirms the relevance of the new calculations. As one can see also in Fig.~\ref{fig:CSop}, another important difference is that the new cross sections are characterized by a dense resonance structure, which is also contrary to the previously published data. This is certainly due to the very large differences between the well depths of the interaction potential. While for the H$_3$O$^+$ -- He system it is about 354.53 \cmmo \citep{ElHanini2019} and for the NH$_3$ -- H$_2$ complex it is 267 \cmmo \citep{Bouhafs2017}, the well depth of the PES in our calculation is 5-7 times larger, about 1887.2 \cmmo.

In order to analyse the cross sections for $p-$H$_3$O$^+$ -- $p$-H$_2$ collision (see Fig.~\ref{fig:CSpp}) we selected only two transition (both with $\Delta j = 0$ ). The solid black line shows our cross sections for the $1_1^- \rightarrow 1_1^+$ inversion transition, while the solid blue line refers to $2_1^+ \rightarrow 2_2^+$. The results are compared with the corresponding data for $p-$H$_3$O$^+$ -- He \citep{ElHanini2019} and $p-$NH$_3$ -- $p-$H$_2$ \citep{Bouhafs2017} collisions. The observed relative differences are significantly smaller for the transition within $j=1$, but still very large for the $j=2$ transition. The strong resonance structures in the cross sections are significant for this type of collisions as well. The differences are mainly due to these resonances in the case of $1_1^- \rightarrow 1_1^+$ transition, while for the $2_1^+ \rightarrow 2_2^+$ process a larger relative difference is observed, from a factor of 2 up to more than an order of magnitude.

\begin{figure}
    \centering
		\includegraphics[width=\columnwidth]{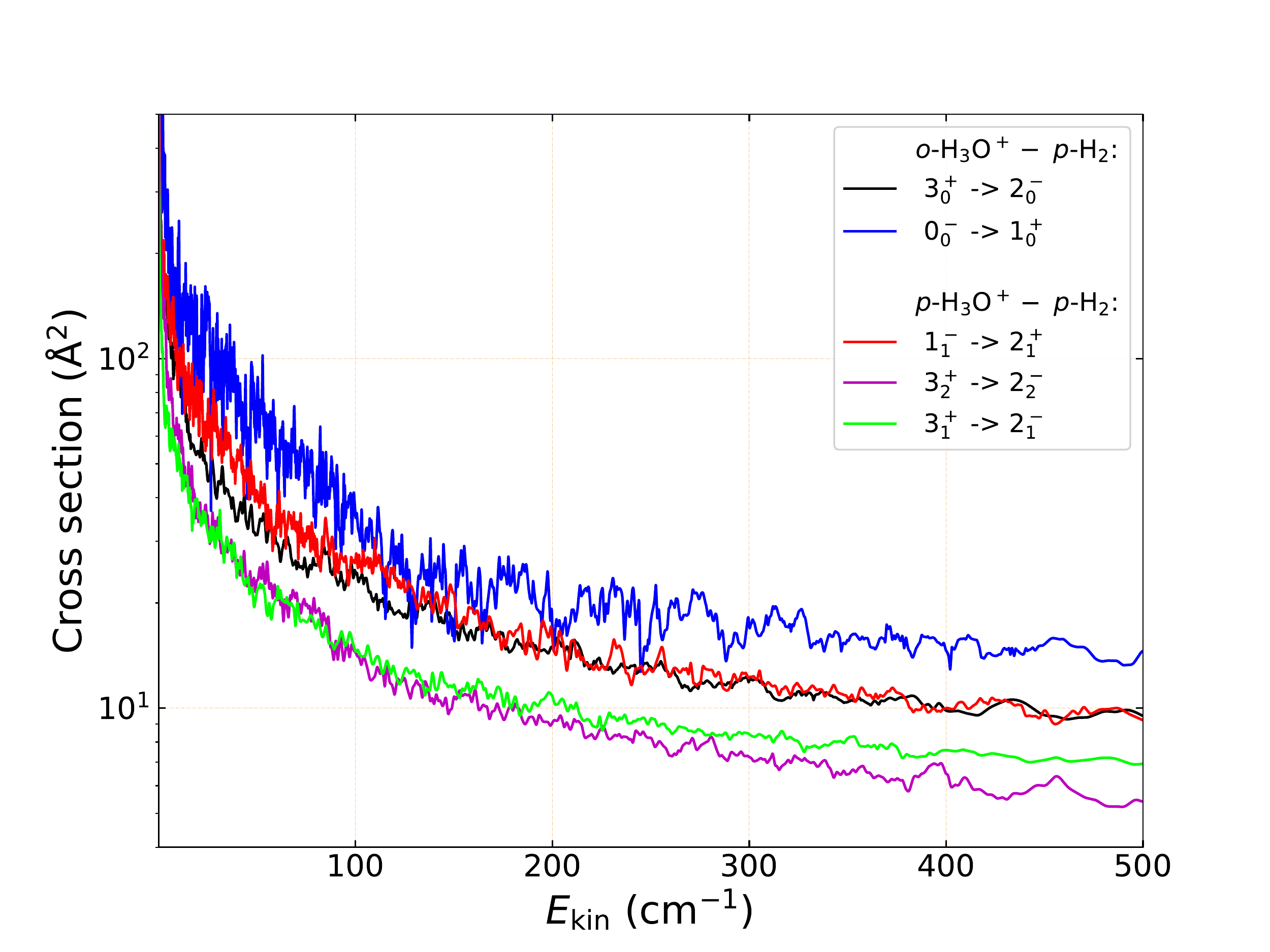}
    \caption{Rotational de-excitation cross sections for the low-frequency (<~1000 GHz) transitions in $o-$H$_3$O$^+$ and $p-$H$_3$O$^+$ collisions with $p-$H$_2$.}
    \label{fig:CSmx}
\end{figure}

In Fig.~\ref{fig:CSmx} we compare the cross sections for all the possible low-frequency transitions ($\nu_{\mathrm{i} \rightarrow \mathrm{f}} < 1000$ GHz) up to 500 \cmmo  collision energies. Transition both for $o-$H$_3$O$^+$ and $p-$H$_3$O$^+$ nuclear spin isomers were considered, for which we also present radiative transfer calculations later (see subsection~\ref{sec:Excit}). The $\sigma$ do not deviate strongly, the maximal difference between them is within a factor of 3. It is not surprising that the largest cross sections are calculated for the case of $0_0^- \rightarrow 1_0^+$ transition into the ground state. As one can see, all $\sigma$ are characterized with a very strong resonance behaviour, which indicates that a
a very dense sampling of collision energy (with a step size $E_\mathrm{step}\sim0.1$ \cmmo) is required to adequately compute the corresponding rate coefficients (especially below 300-400 \cmmo, where the resonances are the strongest).

\subsection{Rate coefficients}\label{sec:RCs}

\begin{figure}
    \centering
        \includegraphics[width=\columnwidth]{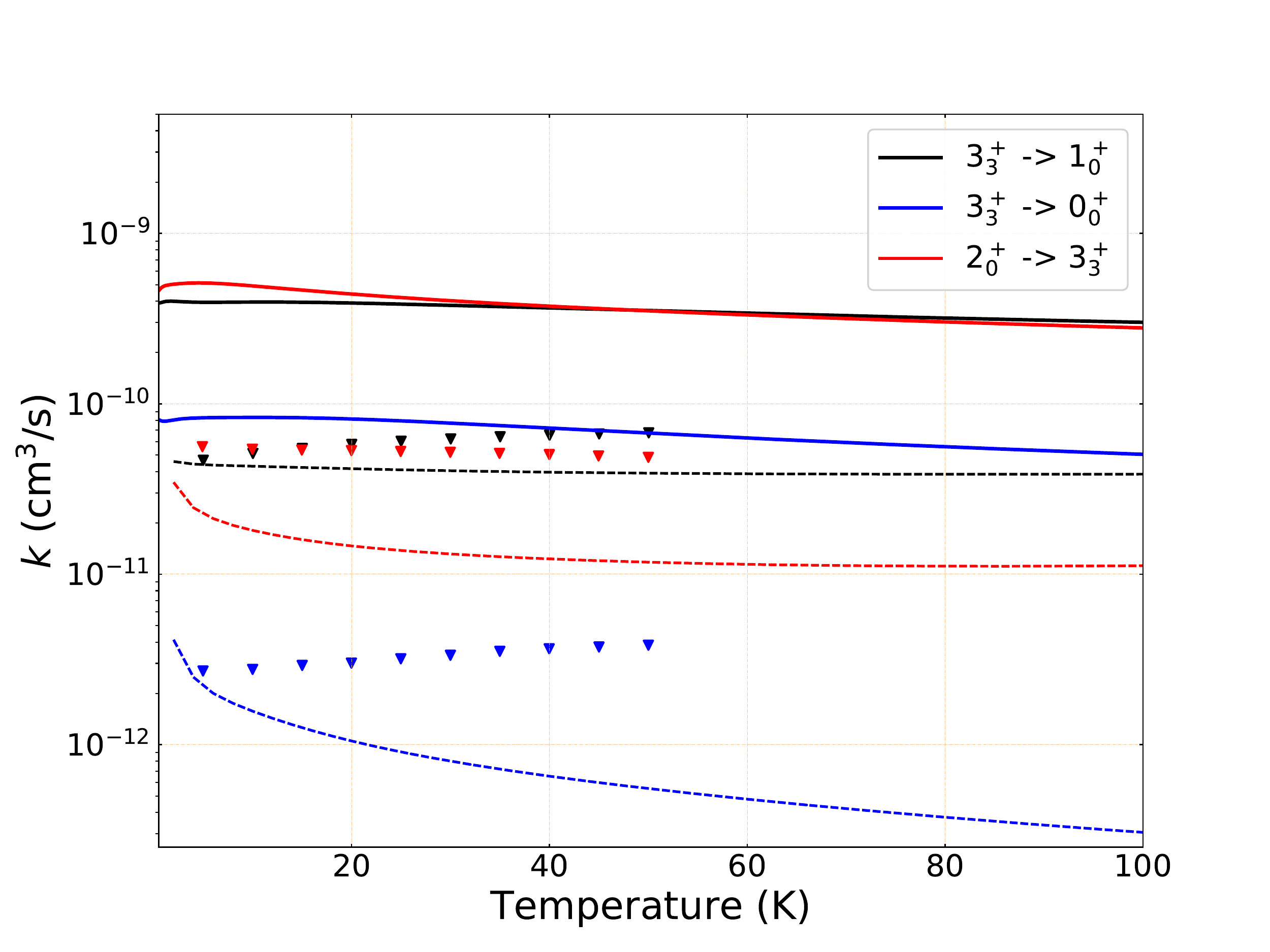}
    \caption{Kinetic temperature dependence of the rate coefficients for some selected rotational transitions in collision of $o-$H$_3$O$^+$ with $p-$H$_2$. Our results (solid lines) are compared with the corresponding data of \citet{Bouhafs2017} ($o-$NH$_3$ -- $p-$H$_2$, dashed lines) and \citet{ElHanini2019} ($o-$H$_3$O$^+$ -- He, triangles). Identical colours indicate identical transitions.}
    \label{fig:RCop}
\end{figure}

Once the state-to-state cross sections were calculated we computed the corresponding rate coefficients for the H$_3$O$^+$ -- H$_2$ collision (see subsection~\ref{sec:ScatCalc} for details). In Fig.~\ref{fig:RCop} the rate coefficients for the $o-$H$_3$O$^+ -$  $p$-H$_2$ collision are compared with the available data from the literature for the same transitions presented in Fig.~\ref{fig:CSop}. The large difference between our results and the collisional rates of both \citet{ElHanini2019} and \citet{Bouhafs2017} is even more visible in this plot in the whole temperature range (i.e. up to $T_\mathrm{kin} = 100$ K). The smaller deviation is observed for the  $3_3^+ \rightarrow 1_0^+$ transition (one order of magnitude compared to two orders of magnitude for the other transition). In general the relative difference between the corresponding $k(T)$ is smaller for higher absolute rate values. It is worth noting also that there is no direct scaling (i.e. linear dependence) observed between the rate coefficients compared in Fig.~\ref{fig:RCop} with respect to kinetic temperature change.

\begin{figure}
    \centering
        \includegraphics[width=\columnwidth]{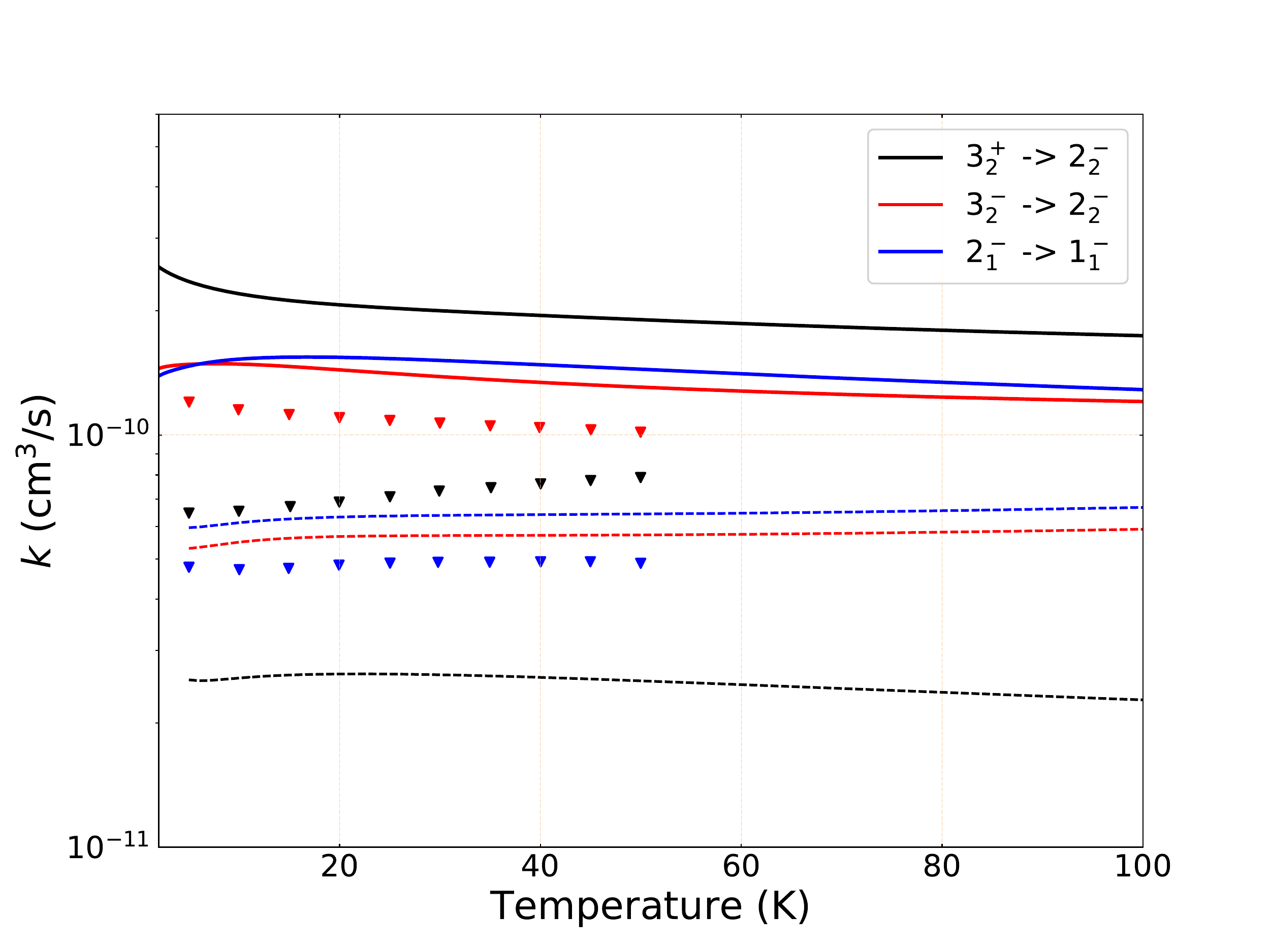}
    \caption{Kinetic temperature dependence of the rate coefficients for some selected rotational transitions in collision of $p-$H$_3$O$^+$ with $p-$H$_2$. Our results (solid lines) are compared with the corresponding data of \citet{Bouhafs2017} ($p-$NH$_3$ -- $p-$H$_2$, dashed lines) and \citet{ElHanini2019} ($p-$H$_3$O$^+$ -- He, triangles). Identical colours indicate identical transitions.}
    \label{fig:RCpp}
\end{figure}

As it is expected, the relative difference between the calculated $k(T)$ for $p-$H$_3$O$^+ -$  $p$-H$_2$ collision and the corresponding data from the literature \citep{Bouhafs2017,ElHanini2019} is similar to those for the $o-$H$_3$O$^+ -$  $p$-H$_2$ process. Fig.~\ref{fig:RCpp} shows the temperature dependence of the $p-$H$_3$O$^+ -$  $p$-H$_2$ thermal rate coefficients for some selected transitions, compared with the previously published data. It is worth to mention however that we selected different states from those shown in Fig.~\ref{fig:CSpp}, because of the lack of collisional rate coefficients in the literature for those transitions. The magnitude of deviation does not depend clearly on the absolute values of the rates, however it is rather small (from about 20\% up to about factor of 2) in the case of the $3_2^- \rightarrow 2_2^-$ transition, which is the lowest by the amplitude in our calculations. While our results show a rather strong resemblance for the $3_2^- \rightarrow 2_2^-$ and $2_1^- \rightarrow 1_1^-$ transitions, this is not the case with corresponding data of \citet{ElHanini2019}. The rate coefficients for these transitions in the work of \citet{Bouhafs2017} also show a strong correlation, but our results are significantly larger (by about a factor of 2). The largest differences (up to an order of magnitude) are found for the $3_2^+ \rightarrow 2_2^-$ de-excitation process, while compared with the results of \citet{Bouhafs2017} for NH$_3$.

\begin{figure}
    \centering
        \includegraphics[width=\columnwidth]{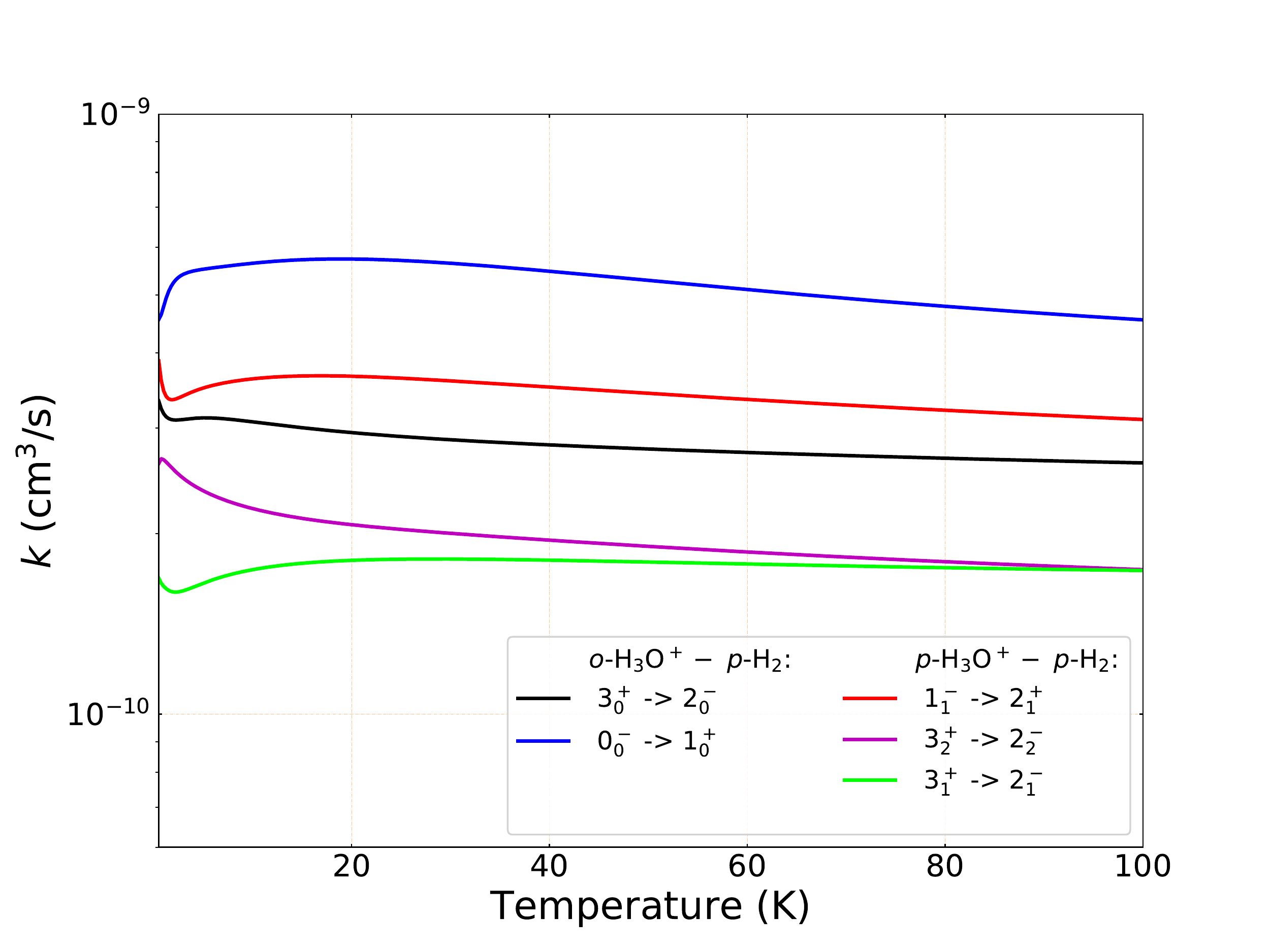}
    \caption{Kinetic temperature dependence of the rate coefficients for the low-frequency ($\nu<1000$ GHz) transitions in both $o-$H$_3$O$^+$ and $p-$H$_3$O$^+$ collisions with $p-$H$_2$.}
    \label{fig:RCmx}
\end{figure}

Fig.~\ref{fig:RCmx} compares the de-excitation rates for the same low-frequency transitions studied in Fig.~\ref{fig:CSmx}. Both the $o-$H$_3$O$^+$ and $p-$H$_3$O$^+$ nuclear spin isomers were considered again. One can see that the absolute difference between the rates varies from some percents up to a factor of 3. It is worth to mention again that the dependence of the rate coefficients on the temperature is not uniform, which is especially significant at lower $T_\mathrm{kin}$. Above 60 K the rate coefficients decrease rather monotonically, and a constant scaling factor between them could be easily derived for this temperature region. 

After analyzing all the state-to-state rate coefficients some important conclusion could be drawn. In the 10 to 100 K interval the temperature dependence is not strong, which is in accordance with the Langevin capture model. Above 40-50 K all the rate coefficients are monotonically decreasing with temperature, which allows one to use simple extrapolation methods towards higher kinetic temperatures, however the precision of such extrapolations are questionable.

\begin{figure}
    \centering
        \includegraphics[width=\columnwidth]{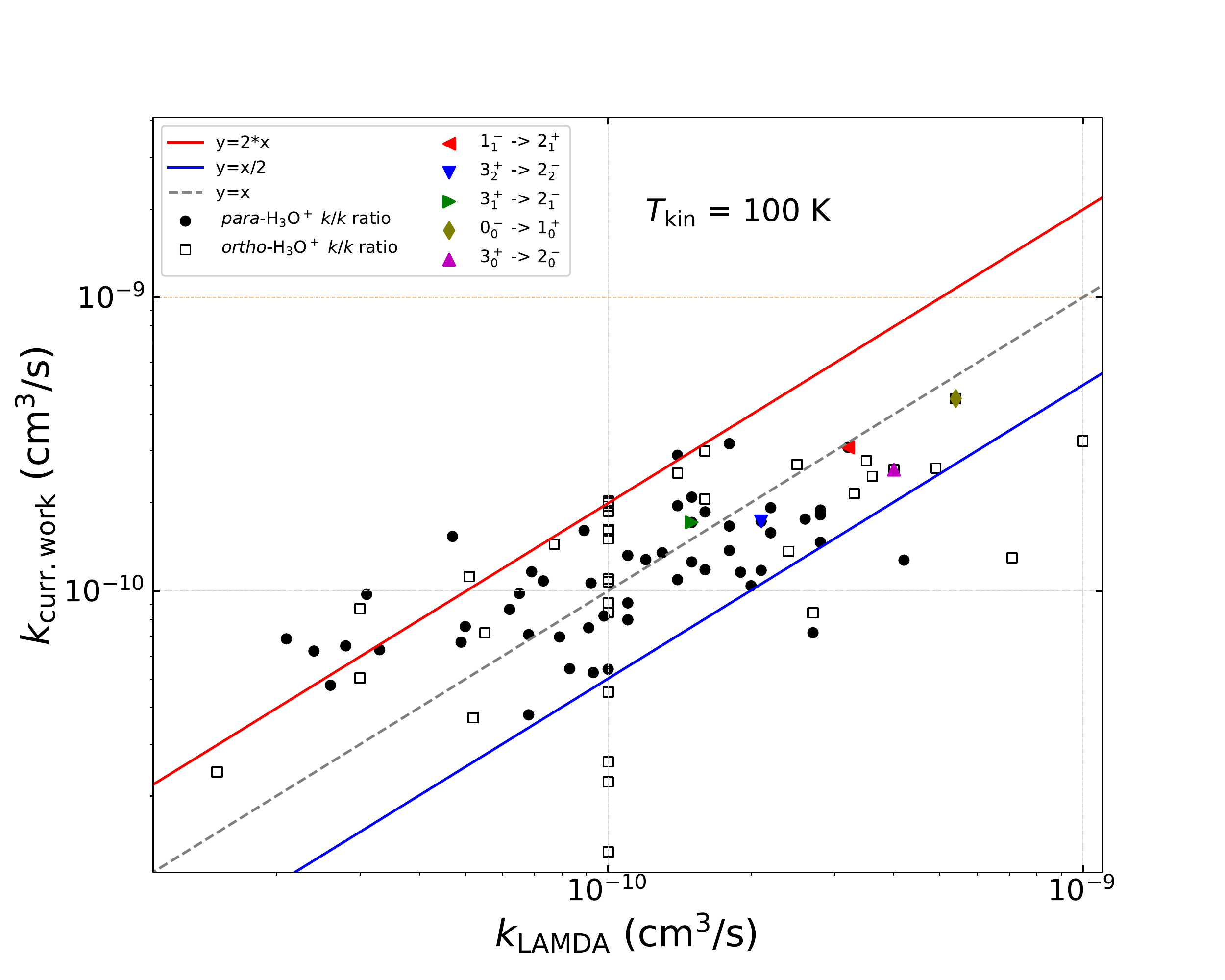}
    \caption{State-to-state ratios of our recent thermal rate coefficients and the corresponding data listed in the \texttt{LAMDA} database \citep{Schoier2005}. Collisional rates both for $o-$H$_3$O$^+$ and $p-$H$_3$O$^+$ are considered at 100~K kinetic temperature.}
    \label{fig:RCcomp}
\end{figure}

In order to demonstrate the importance of new rate coefficients in the potential astrophysical applications we compare our results with the most recent $k(T)$ listed in the \texttt{LAMDA} database \citep{Schoier2005}. Currently these rate coefficients are used for the radiative transfer modelling, which involves hydronium. The state-to-state ratios of the collisional rates both for $o-$H$_3$O$^+$ and $p-$H$_3$O$^+$ are shown in Fig.~\ref{fig:RCcomp} at 100 K kinetic temperature. As one can see the $k(T)$ do not correlate well, i.e. no linear scaling from the data of \texttt{LAMDA} database to our new rates can be found. In general the ratio varies between factors of 0.5 and 2 (except for some $o-$H$_3$O$^+$-transitions) and no significant difference can be found for $ortho$ and $para$ nuclear symmetries. For most of the transition the following tendency is valid: as the magnitude of the $k(T)$ is increasing their relative ratio is decreasing. In comparison with the data from the literature our rates are higher for the low-magnitude transitions (up to about $2 \times 10^{-10}$ cm$^3/$s), while for higher rates they are lower. We would like to emphasise again that due to this non-linearity the new collisional rate coefficients are important for adequate modelling of hydronium in astrophysical media.

The corresponding ratios for the low-frequency transitions (highlighted with alternative colours) are indicated at 50 K as well. While the the \texttt{LAMDA} database currently contains only collisional data at 100 K, it is important to provide temperature-dependent rates as well, since the rates can vary (decrease) with increasing $T_\mathrm{kin}$.

\subsection{Excitation of H$_3$O$^+$ in low-temperature interstellar regions}\label{sec:Excit}

\begin{figure*}
    \centering
		\includegraphics[width=\textwidth]{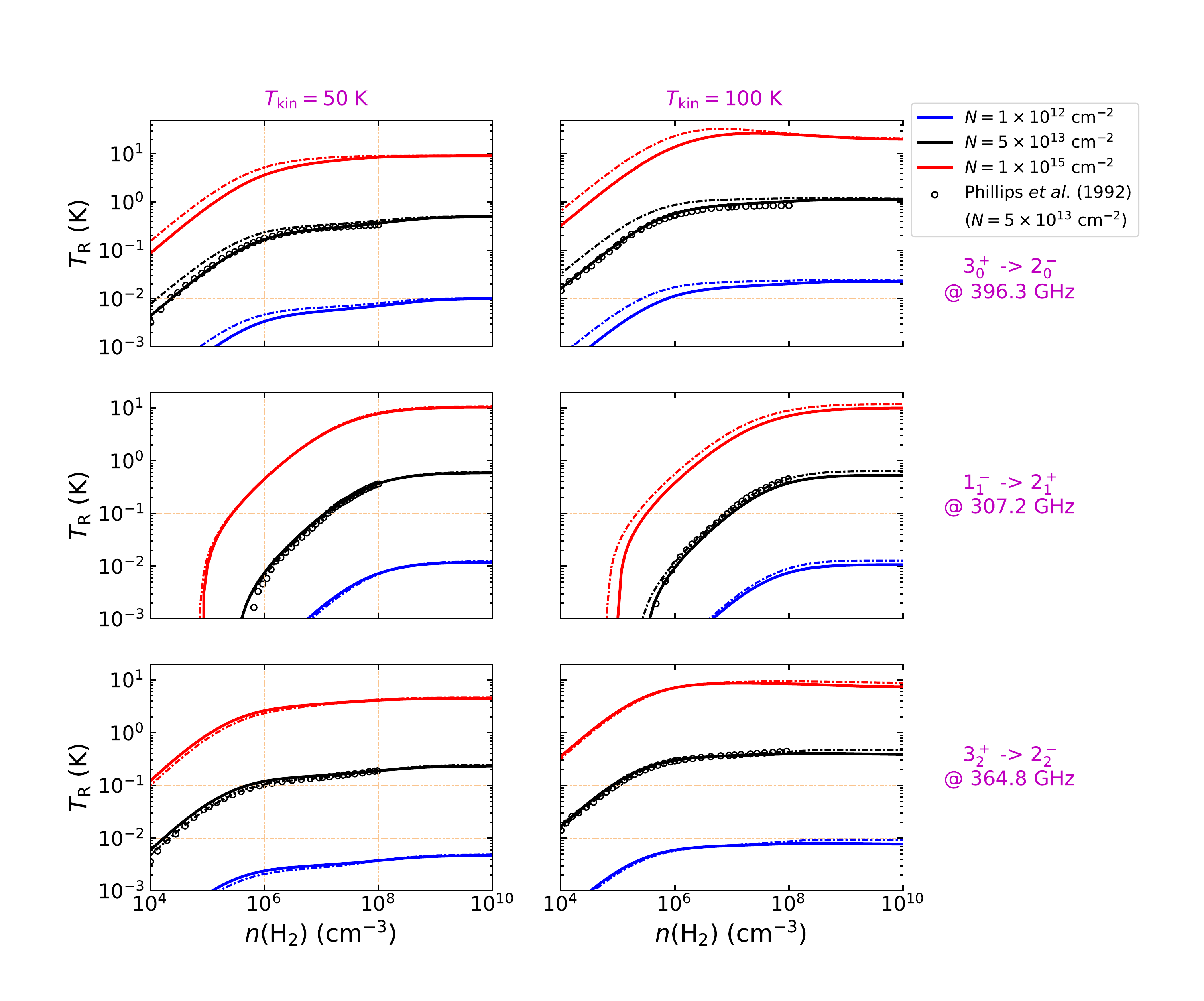}
    \caption{Radiance temperatures $T_\mathrm{R}$ for the H$_3$O$^+$ cation's  307, 365 and 396 GHz lines as a function of molecular density $n$(H$_2$), computed using the rate coefficients from the \texttt{LAMDA} database \citep[dash-dot lines]{Schoier2005} and those from this paper (solid lines). The 3 panels on the left side represent the results at 50 K kinetic temperature for the 3 transitions studied, while on the right side panels the results at 100 K are shown. The different colours of the subplots denote different column densities, ranging from $N=10^{12}$ up to $N=10^{15}$ cm$^{-2}$.}
    \label{fig:TRvsTkin}
\end{figure*}

In order to estimate the impact of our new collisional data on the astrophysical observables, we have computed the excitation of both $ortho$- and $para$-H$_3$O$^+$ using the large velocity gradient \texttt{RADEX} non-LTE radiative transfer computer program (see subsection \ref{sec:RadTRCalc} for details). We have used the rate coefficients computed for collision with $p$-H$_2$, presented in the previous section. For comparison, we provided the same radiative transfer calculations using the corresponding $k(T)$ data from the \texttt{LAMDA} database \citep{Schoier2005}. The radiance (brightness) temperatures for the $1_1^- \rightarrow 2_1^+$, $3_2^+ \rightarrow 2_2^-$ and $3_0^+ \rightarrow 2_0^-$ transitions are compared with results of \citet{Phillips1992} (see Fig.~\ref{fig:TRvsTkin}). The latter de-excitation processes are characterized with 307.2, 364.8 and 396.3 GHz transition frequencies, respectively (see Fig.~\ref{fig:CSmx} for the corresponding cross sections and Fig.~\ref{fig:RCmx} for the $k(T)$).

First we analysed the optical depth ($\tau$) for these transitions at $N=10^{15}$~cm$^{-2}$ column density. We found that all the 4 transitions are optically "thin", since we found very low $\tau$ values for them under strict LTE-conditions (above~$10^{9}$ cm$^{-3}$ hydrogen densities):

\begin{itemize}
  \item $\tau \approx 0.2-0.22$  for the $3_0^+ \rightarrow 2_0^-$ transition (396.3 GHz),
  \item $\tau \approx 0.15-0.17$  for the $1_1^- \rightarrow 2_1^+$ transition (307.2 GHz),
  \item $\tau \approx 0.08-0.1$  for the $3_2^+ \rightarrow 2_2^-$ transition (364.8 GHz),

\end{itemize}
where the two $\tau$-values refer to results obtained using our new $k(T)$ data and those of from the \texttt{LAMDA} database \citep{Schoier2005}, respectively.
At lower H$_2$ densities (below~$10^{6}$ cm$^{-3}$) the calculated optical depth values are lower $(<0.1)$, except of the $1_1^- \rightarrow 2_1^+$ (307.2 GHz) transition, for which $\tau \approx 1$ between $10^{3}-10^{4}$ cm$^{-3}$ H$_2$ densities.

Fig.~\ref{fig:TRvsTkin} shows the dependence of $T_\mathrm{R}$ radiance temperature on $n$(H$_2$) gas density for the above mentioned transitions. We analysed this dependence for different column densities $N$ ranging from $1\times10^{12}$ up to $1\times10^{15}$ cm$^{-2}$ at 50 and 100 K gas temperatures. As one can see there is a good qualitative agreement between the calculated radiance temperatures using our new rate coefficients and those listed by \citet{Phillips1992} for the whole molecular density range. However, while analysing the results obtained using the \texttt{LAMDA} rate coefficients we observe some deviations. These deviations with respect to the results obtained with our $k(T)$ data are ranging from some percents up to a factor of 2 (compare the solid and dash-dot curves on Fig.~\ref{fig:TRvsTkin}). As one can expect, the deviations are more significant under non-LTE conditions, i.e. below $10^{6}$ cm$^{-3}$ hydrogen densities, where they strongly depend both on $T_\mathrm{kin}$ as well as on the type of the transition. It is worth to mention however that at 100 K slight differences (up to about 10 \%) are observed also at high H$_2$ densities, where the collisional rates do not contribute in the modelling. These differences can be explained by the different number of rotational leveles and transitions involved in the two set of radiative transfer calculations. Thus, in the modelling with our rate coefficients the collisional data for the lowest 20 and 11 rotational states are included for the $para$- and $ortho$-H$_3$O$^+$, respectively. At the same time, in the modelling with the \texttt{LAMDA} rate coefficients the corresponding data are available only for the lowest 14 and 9 rotational levels, respectively. As one can see, at lower temperatures (i.e. at 50 K) these differences do not lead to large discrepancies between the two calculations, however at 100 K the additional rotational states have significant impact on the results. We found that the disagreement between the two calculated set of $T_\mathrm{R}$ is the largest in the case of $3_0^+ \rightarrow 2_0^-$ $o$-H$_3$O$^+$ -- $p$-H$_2$ transition and is stronger at higher kinetic temperature. It does not depend however on the column density. The observed threshold (i.e. the critical density) to reach LTE conditions is about $10^{8}$ cm$^{-3}$ H$_2$ density for all considered de-excitation processes.

\begin{figure*}
    \centering
		\includegraphics[width=\textwidth]{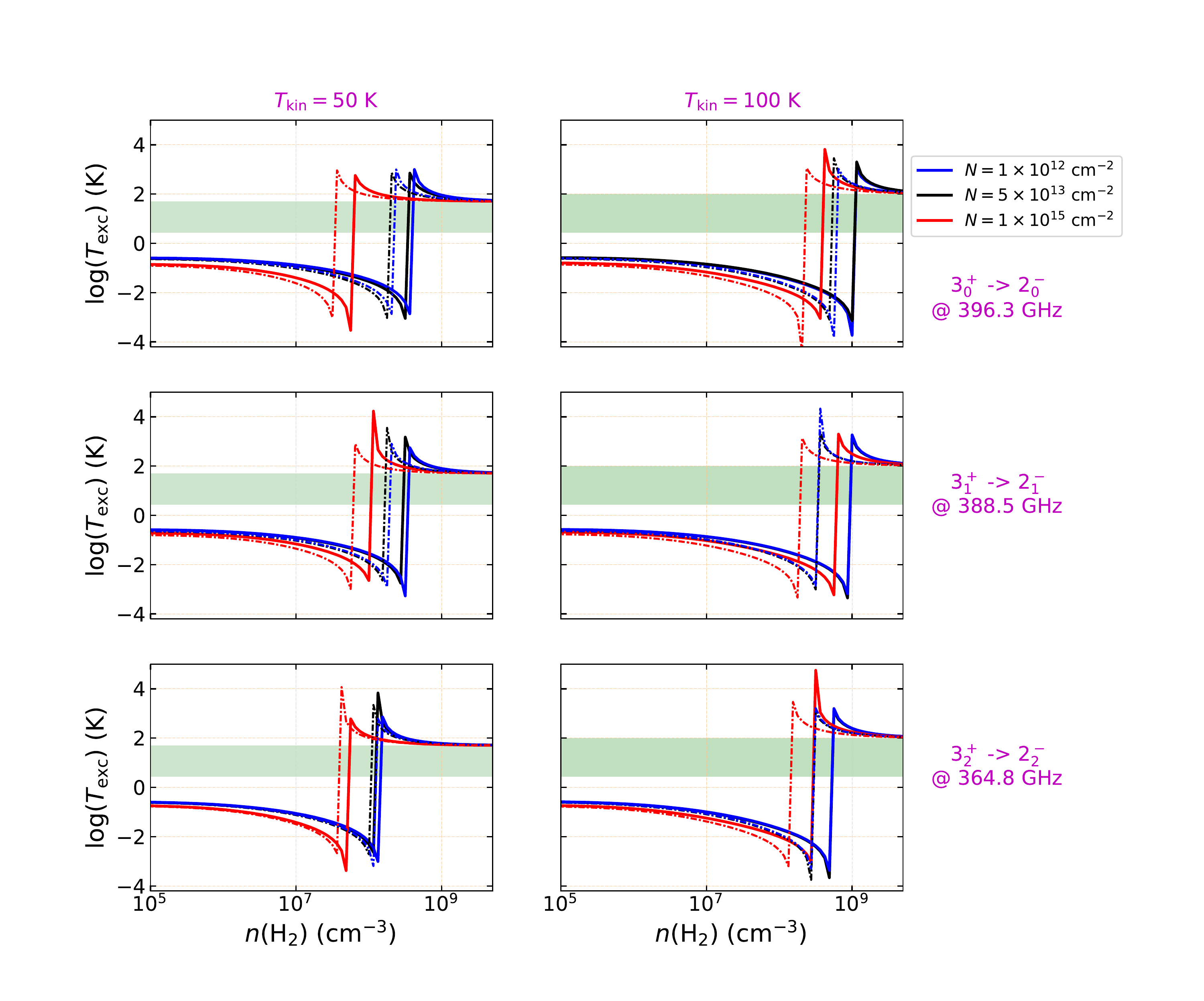}
    \caption{Logarithm of the excitation temperatures $\log(T_\mathrm{exc})$ for the H$_3$O$^+$ cation's 365, 389 and 396 GHz emission lines as a function of molecular density~$n$(H$_2$), computed using the rate coefficients from the \texttt{LAMDA} database \citep{Schoier2005} (dash-dot lines) and those from this paper (solid lines). The  values at the ordinate are multiplied by a factor $\varepsilon=\pm1$, the sign of  which corresponds to the sign of $T_\mathrm{exc}$. The green horizontal area defines the interval of temperatures ranging from $T_\mathrm{bg}=2.7$ K up to $T_\mathrm{kin}$. The definition and parameters used in the individual panels are identical to those of in Fig.~\ref{fig:TRvsTkin}.}
    \label{fig:TexcRvsTkin}
\end{figure*}

To further studying the effect of the new rate coefficients on the radiative transfer models we provide also a comparative analysis for the $T_\mathrm{exc}$ excitation temperatures. In Fig.~\ref{fig:TexcRvsTkin} one can see the dependence of $T_\mathrm{exc}$ on H$_2$ molecular density in case of some low-frequency transitions, for which population inversion effects were observed. In particular the $3_0^+ \rightarrow 2_0^-$, $3_1^+ \rightarrow 2_1^-$ and $3_2^+ \rightarrow 2_2^-$ de-excitation processes were considered with 396.3, 388.5 and 364.8 GHz transition frequencies, respectively. In the subplots of Fig.~\ref{fig:TexcRvsTkin} we compared again two set of results, which are obtained using our new $k(T)$ and those from the \texttt{LAMDA} database. Analogously to the previous case, we have analysed the results for column densities between $1\times10^{12}$ and $1\times10^{15}$ cm$^{-2}$ at 50 and 100 K kinetic temperatures. Unfortunately, for these quantities we could not find any available data in the literature. For better visualization the plots show the logarithm of the excitation temperatures multiplied by by $\varepsilon$, where $\varepsilon=1$ for positive and $\varepsilon=-1$ for negative $T_\mathrm{exc}$ values, respectively. As one can see for all 3 transitions under consideration a population inversion is observed in a very broad gas density range, starting from low $n($H$_2)$ densities (below $10^{4}$ cm$^{-3}$). The width of this region strongly depends both on the column density and the kinetic temperature. Between $10^{8}-10^{9}$ cm$^{-3}$ however, i.e. around the threshold to reach strict LTE-conditions, the population inversion is quenched, and at high densities there are almost no differences between the two sets of data. On the other hand, while we compare the results for the particular transitions under the same physical conditions, which were obtained with the two set of rates (see the solid and dash-dot curves in Fig.~\ref{fig:TexcRvsTkin}), we find significant differences. First, there is a rather large shift in the critical density region, where $T_\mathrm{exc}$ change sign from $"-"$ to $"+"$. This shift is varying between some percents and about a factor of 2. Slight deviations are also observed in the amplitudes of excitation temperatures between the two set of data near these turning points. These deviations are usually higher in case of larger column densities and lower kinetic temperatures, for which we observe differences as high as an order of magnitude (see for example the $3_1^+ \rightarrow 2_1^-$ and $3_2^+ \rightarrow 2_2^-$ transitions at 50 K). It is also important to note that the minimal and maximal amplitudes in the critical density region are very close for all of the studied transitions and does not depend strongly on the particular physical conditions. Aalto and coworkers showed earlier \citep{Aalto2011} that the results of radiative transfer calculations are also sensitive to the adopted background temperature. So for a reliable interpretation of observations the $T_\mathrm{bg}$ value should be chosen carefully.

\section{Conclusions}\label{sec:Conclusions}

We have presented state-to-state rotational de-excitation cross sections and kinetic rate coefficients for collisions of $ortho-$H$_3$O$^+$ and $para-$H$_3$O$^+$ with $p-$H$_2$, which is one of the most abundant colliders in interstellar clouds. The theoretical calculations are performed within an accurate close-coupling formalism. Our recent, explicitly correlated five-dimensional rigid-rotor potential energy surface \citep{Demes2020} was used for collisional dynamics studies. A broad kinetic energy range was considered ($<800$ \cmmo), which allowed one to calculate the corresponding state-to-state rate coefficients up to 100~K temperature. Since accurate data for the studied collisional system is missing in the literature, we compare our results with the corresponding data for the isoelectronic H$_3$O$^+$ -- He \citep{ElHanini2019} and NH$_3$ -- H$_2$ \citep{Bouhafs2017} collisions. In most cases significant differences are observed (as high as an order of magnitude) both for the cross sections and rate coefficients, where our new data are notably larger than the corresponding data from the literature in case of the analysed de-excitation processes. We also compared the rate coefficients for the individual rotational transitions with the most recent data provided in the \texttt{LAMDA} database \citep{Schoier2005}. The deviations are rather large, mainly within a factor of 4. It is also worth to mention that no general, linear scaling rule could be established to find a correspondance between the literature data and those of our recent work. Based on these findings we can conclude that the new collisional rates are obviously important and allows one a more adequate modelling of hydronium observations in astrophysical media.

In order to investigate the effect of the new rate coefficients on the excitation of $ortho$-H$_3$O$^+$ and $para$-H$_3$O$^+$, we have computed the excitation of these species using a non-LTE radiative transfer code under various physical conditions that are typical of molecular clouds. As the comparison with the results obtained from \texttt{LAMDA} rate coefficients has shown, our new collisional data have a significant, non-negligible impact on the astrophysical observables in such models. This effect on the excitation of hydronium is more remarkable at lower hydrogen densities, where the LTE-conditions are not fulfilled. It is worth to mention however that some discrepancies for the radiance temperatures were found also under strict LTE-conditions. These discrepancies are due to the fact that our new collisional data involve more rotational levels compared to the \texttt{LAMDA} database, which can significantly contribute to the radiative transfer models at higher temperatures. The column densities of H$_3$O$^+$ in interstellar clouds thus should be revisited based on the new collisional data. This can be used then as an indirect way for more precise estimates of the gas-phase formation of O$_2$ and H$_2$O molecules in the ISM, since these species primarily formed by dissociative recombination of hydronium cations.

It is also important to note that the calculated rate coefficients slightly depend on the kinetic temperature. Nevertheless, for more precise astrophysical applications, especially to study warmer astronomical regions (with temperatures around $250-300$ K), it is required to extend the collisional energy and kinetic temperature range. For this reason, our future studies will be addressed to the excitation of H$_3$O$^+$ at higher temperatures, considering rotational cross sections up to 1500-1600 \cmmo collision energies. In addition, since it is crucial in the high-temperature regions of the ISM, we will extend our calculations also for collisional processes involving $ortho$-H$_2$.

\section*{Acknowledgements}

We acknowledge financial support from the European Research Council (Consolidator Grant COLLEXISM, Grant Agreement No. 811363) and the Programme National “Physique et Chimie du Milieu Interstellaire” (PCMI) of CNRS/INSU with INC/INP cofunded by CEA and CNES. We wish to acknowledge the support from the CINES/GENCI for awarding us access to the OCCIGEN supercomputer within the A0070411036 project as well as the GRICAD infrastructure (https://gricad.univ-grenoble-alpes.fr), which is supported by Grenoble research communities. F.L. acknowledges the Institut Universitaire de France.

\section*{Supplementary material}

The \texttt{RADEX}-compatible molecular data files, which are used in the radiative transfer calculations, and which include the particular state-to-state rate coefficient data for $ortho-$ and $para-$H$_3$O$^+$ collisions with $para-$H$_2$ up to 100 K are provided as supplementary material.

\section*{Data Availability Statement}

The data underlying this article will be shared on reasonable request to the corresponding author.




\bibliographystyle{mnras}
\bibliography{references} 



\bsp	
\label{lastpage}
\end{document}